\documentclass{article}
\usepackage[T1]{fontenc} 
\usepackage[utf8]{inputenc} 
\usepackage{ismir,amsmath,cite,url}
\usepackage{graphicx}
\usepackage{xcolor}

\usepackage{breqn} 
\usepackage{float}    

\title{Evolution of the Informational Complexity of Contemporary Western Music}

\oneauthor
  {Thomas Parmer, Yong-Yeol Ahn} {School of Informatics, Computing, and Engineering \\ Indiana University, Bloomington, IN, USA \\ {\tt tjparmer@indiana.edu, yyahn@iu.edu}}

\begin{document}

\maketitle
\begin{abstract}

We measure the complexity of songs in the Million Song Dataset (MSD) in terms of pitch, timbre, loudness, and rhythm to investigate their evolution from 1960 to 2010. By comparing the Billboard Hot 100 with random samples, we find that the complexity of popular songs tends to be more narrowly distributed around the mean, supporting the idea of an inverted U-shaped relationship between complexity and hedonistic value. We then examine the temporal evolution of complexity, reporting consistent changes across decades, such as a decrease in average loudness complexity since the 1960s, and an increase in timbre complexity overall but not for popular songs. We also show, in contrast to claims that popular songs sound more alike over time, that they are not more similar than they were 50 years ago in terms of pitch or rhythm, although similarity in timbre shows distinctive patterns across eras and similarity in loudness has been increasing.
Finally, we show that musical genres can be differentiated by their distinctive complexity profiles.

\end{abstract}
\section{Introduction}\label{sec:introduction}

Our everyday life is surrounded by cultural products; we wake up to a song, read a book on the subway, watch a movie with friends, or even travel far to admire a piece of art. Despite such pervasiveness, we cannot fully explain why we like a particular song over others or what makes something a great piece of art. Although the perceived quality of a piece is affected by numerous contextual factors, including one's cultural, social, and emotional background, theories suggest that preference, or `hedonistic value', may also be affected by innate properties of the products, such as \emph{novelty} and \emph{complexity}~\cite{sigaki2018history,Abdallah,narmour1990analysis}.  In particular, a popular theory suggests there is a \emph{Goldilocks principle} --- that just the right amount of novelty or complexity elicits the largest amount of pleasure, whereas pieces with too little or too much complexity are less interesting and enjoyable~\cite{berlyne1970novelty}. On the other hand, cultural products are also \emph{fashionable} --- what is popular now may be completely out of fashion next month. Such seemingly contrasting observations prompt us to ask the following questions: \emph{as fads come and go, is there still a consistent preference towards the optimal amount of complexity in cultural products? How has the complexity of contemporary cultural products changed over time?}

This question may apply to any type of cultural product, but we focus here on the complexity of contemporary Western songs. Although various studies have already reported evidence of the `inverted U-shaped' relationship between perceived complexity and the pleasantness of music in terms of individual-level preference~\cite{Abdallah, streich2006music, heyduk1975rated}, evidence of this preference at the population-level is unclear~\cite{simonton1994computer, eerola2000cognitive, parry2004musical}, and many past studies have been limited by the size or extent of the data, in terms of genres or temporal range.

Recently, datasets such as the Million Song Dataset (MSD) began to allow researchers to systematically analyze patterns in music at a massive scale~\cite{Foster2,Dieleman,Serra}.  For example, Serra~\emph{et al.} used musical `codewords' based on song segments in the MSD to identify changes in pitch, timbre, and loudness over time, finding that newer songs restrict pitch transitions, homogenize timbre, and increase loudness (without increasing the variability in loudness)~\cite{Serra}. Mauch~\emph{et al.} used a corpus of 17,000 songs from the Billboard Hot 100 to analyze how popular music has evolved between 1960 and 2010 in the United States; using timbral and harmonic features derived from songs on the Hot 100, they identified three stylistic revolutions that occurred in 1964, 1983, and 1991~\cite{Mauch}.

In this paper, we analyze the large-scale evolution of complexity in contemporary music in terms of pitch, loudness, timbre, and rhythm, during the period from 1960 to 2010 using the Million Song Dataset (MSD)~\cite{Thierry}.  We find that complexity does seem to constrain popularity, as evidenced by the most popular songs (those on the Billboard Hot 100) clustering around average values and exhibiting smaller variance compared to a random sample. However, complexity values do fluctuate over time, as long-term trends are seen in loudness, timbre, and rhythm complexity and in the similarity between songs on the Billboard Hot 100.
Finally, we compare the complexity of different genres and find that genres have characteristic complexity profiles, leading us to hypothesize that complexity may be a factor in an individual's musical selection.

\section{Methods}\label{sec:methods}


\subsection{Data}\label{subsec:data}

The MSD is a dataset of one million songs created by Columbia University's LabROSA in collaboration with The Echo Nest \cite{Thierry}. Each song in the dataset is divided into small temporal segments (based on note onsets) with detailed data derived from the song's audio signal, and includes metadata such as title, artist, year, duration, and genre terms.

Prior to analysis, we filtered the MSD to remove duplicates, songs with missing genre or duration metadata, and songs likely to be commentary pieces (whose title included the tokens `interview', `commentary', `introduction', `discuss', `conference', or `intro'), resulting in a dataset of 905,896 songs.  Some songs also did not have all data types --- pitch, loudness, timbre, rhythm, or year --- that we examine here and were thus left out of the corresponding calculations.  
Due to a limited amount of data from the early years, we restricted our analysis to the period from 1960 to 2010.  The genre of each song was determined by the term (Echo Nest tag) with the strongest weight, although we note that terms are assigned at the artist level so all songs by the same artist are grouped into the same genre.

\begin{figure}
 \centerline{
 \includegraphics[width=\columnwidth]{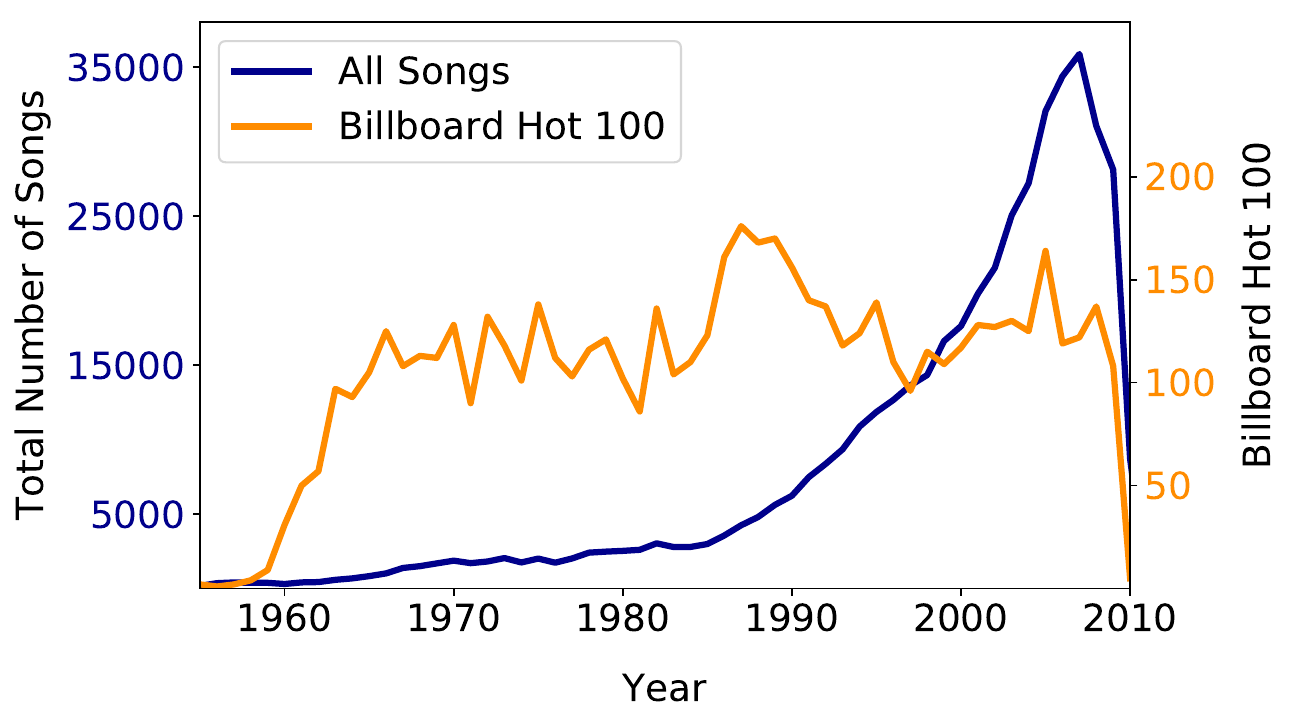}}
 \caption{The number of songs per year.  `All Songs' refers to the filtered MSD dataset, while `Billboard Hot 100' refers to those songs whose title and artist we matched with songs on the Hot 100 as identified in \cite{Mauch}.}
 \label{fig:data_per_year}
\end{figure}

To discover the most popular musical pieces in our dataset, we found 6,661 songs which charted on the Billboard Hot 100 as identified in a previous study \cite{Mauch}.  The number of songs per year in our final dataset is shown in Figure \ref{fig:data_per_year}.

\subsection{Codewords}\label{subsec:codewords}

To estimate complexity, we defined ``codewords'' for each song across four dimensions (pitch, loudness, timbre, and rhythm), similar to a previous study~\cite{Serra}.  
Each codeword is based on a segment of the song.  Pitch and timbre codewords are vectors containing the pitches (based on the binary presence of each of 12 pitches in the chromatic scale) and timbres (based on analysis of the audio signal, with 11 components thresholded into three bins) present in the segment. Loudness codewords are equal to the binned maximum decibel value of the segment.  Similarly, rhythm codewords are defined as the number of average sixteenth notes between segments, where the average sixteenth note is based on the time signature.  We then defined a measure of complexity for each feature per song based on the conditional entropy of each type of codeword.\footnote{Code is available at https://github.com/tjparmer/music-complexity.}


\subsection{Measuring Complexity}\label{subsec:codewords}


Although many studies have examined the relationship between the complexity of a piece and the derived pleasure from it\cite{Abdallah, streich2006music,steck1975preference,north1997experimental,north1997liking,orr2005relationship}, there is no universally adopted way to measure the complexity of a song.  Existing definitions of complexity include hierarchical complexity, dynamic complexity, information-theoretic complexity, and cognitive complexity \cite{pressing1999cognitive,shmulevich2000measures,Gregory,Thul}.  Information-theoretic measures are attractive because they capture the surprise inherent in a pattern, such as the notes played in a musical piece.  Theories propose that music can be understood as the kinetics of expectation and surprise, and that composers seek to elicit emotions by fulfilling or denying these expectations \cite{meyer1967music,Abdallah,zivic2013perceptual}.  In particular, Implication-realization (IR) theory posits that open intervals evoke expectations in a listener and the surprises of these expectations may be related to complexity \cite{streich2006music, narmour1990analysis, zivic2013perceptual}.

Information-theoretic measures include Shannon entropy, joint entropy, conditional entropy, compression or algorithmic complexity~\cite{simon,shmulevich2000measures, Meredith,Gregory}, and more complicated techniques such as pairwise predictability between time series, Hidden Markov Models, Normalized Compression Distance, and predictive information rate~\cite{Foster1,Foster2,Abdallah}.  Previous studies have used information-theoretic quantities to estimate perceived complexity, identify piece similarity, derive psycho-acoustic features, and classify genres~\cite{Gregory, shmulevich2000measures, cilibrasi2004algorithmic, scheirer2000perceived, jennings2004variance, simonton1994computer}.

We use conditional entropy as our measure of complexity, dependent on the immediately preceding symbol, as it is known that events during even short preceding intervals are enough to evoke strong expectations in the listener \cite{Huron,Abdallah,zivic2013perceptual}. Other information-theoretic measures are either more complicated (e.g. predictive information measures), can only be approximated in practice (e.g. Kolmogorov or algorithmic complexity), or do not take past information into account (e.g. Shannon entropy). 


Each song was assigned a complexity value for pitch, loudness, timbre, and rhythm, which is equal to the conditional entropy of the feature codewords:

\begin{dmath}\label{complexity}
H(Y|X) = \sum_{x \epsilon X} p(x) H(Y|X=x) = -\sum_{x \epsilon X} p(x) \sum_{y \epsilon Y} p(y|x) \log_2 p(y|x)
\end{dmath}

where $X$ and $Y$ are possible codewords, $p(x)$ is the probability of observing codeword $x$ and $p(y|x)$ is the probability of observing codeword $y$ given the previous codeword $x$.

\section{Results}\label{sec:results}

\subsection{Complexity and Popularity}

\begin{figure*}
 \includegraphics[width=.54\columnwidth]{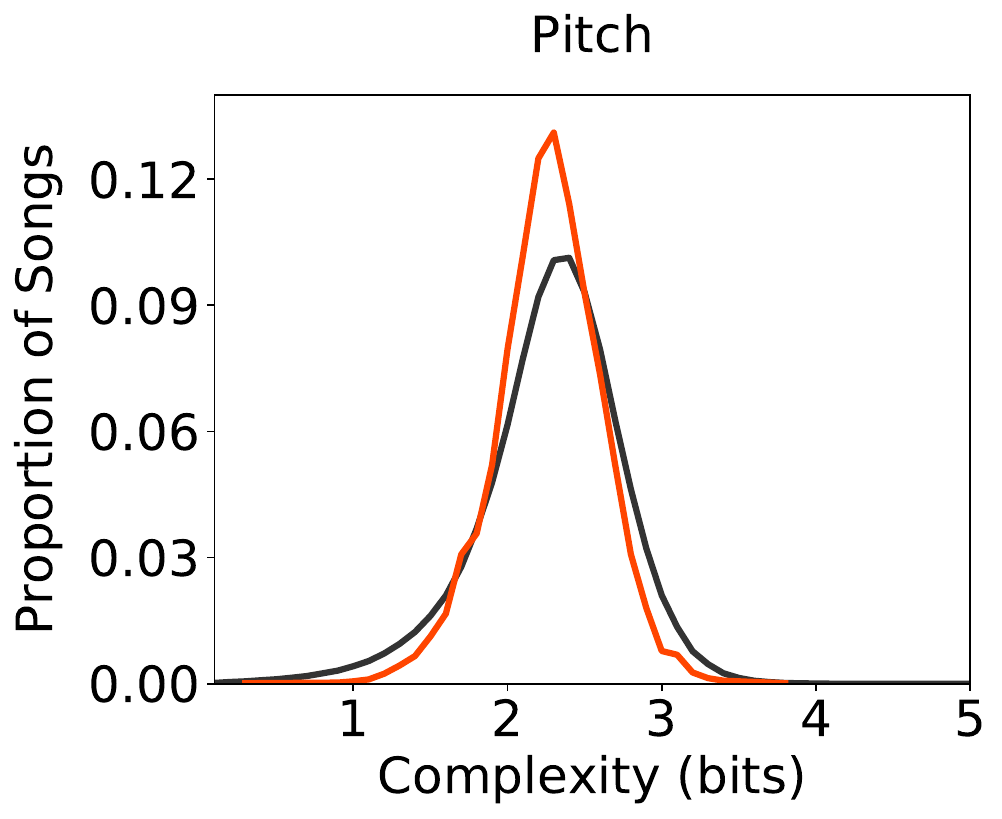}
 \includegraphics[width=.5\columnwidth]{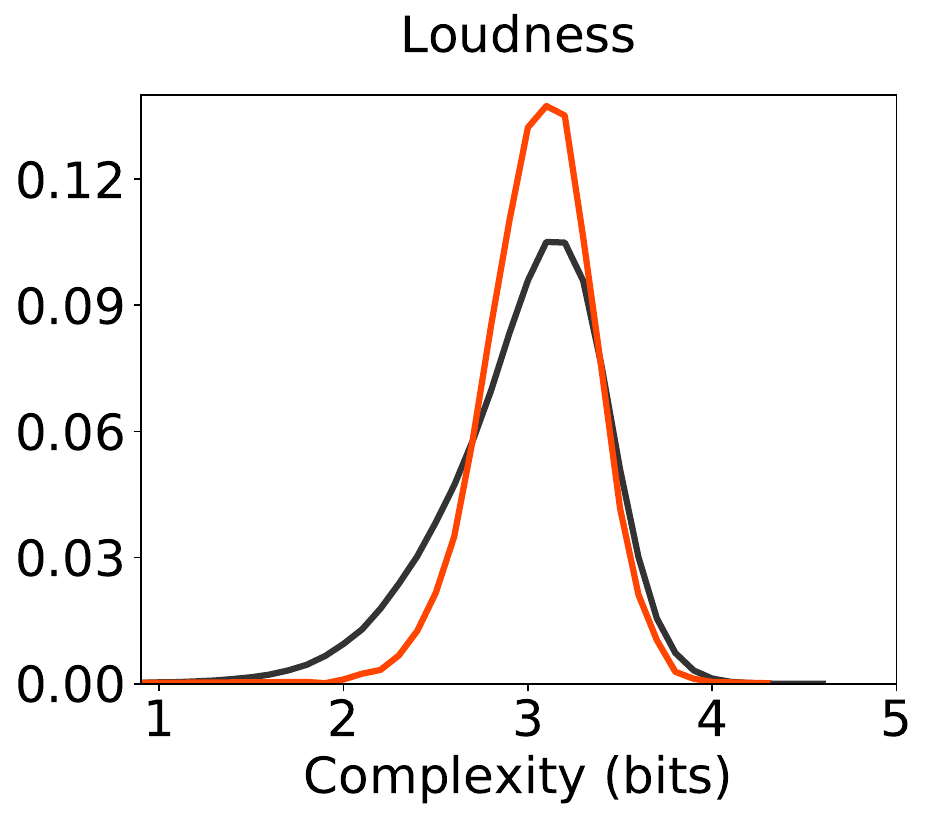}
  \includegraphics[width=.5\columnwidth]{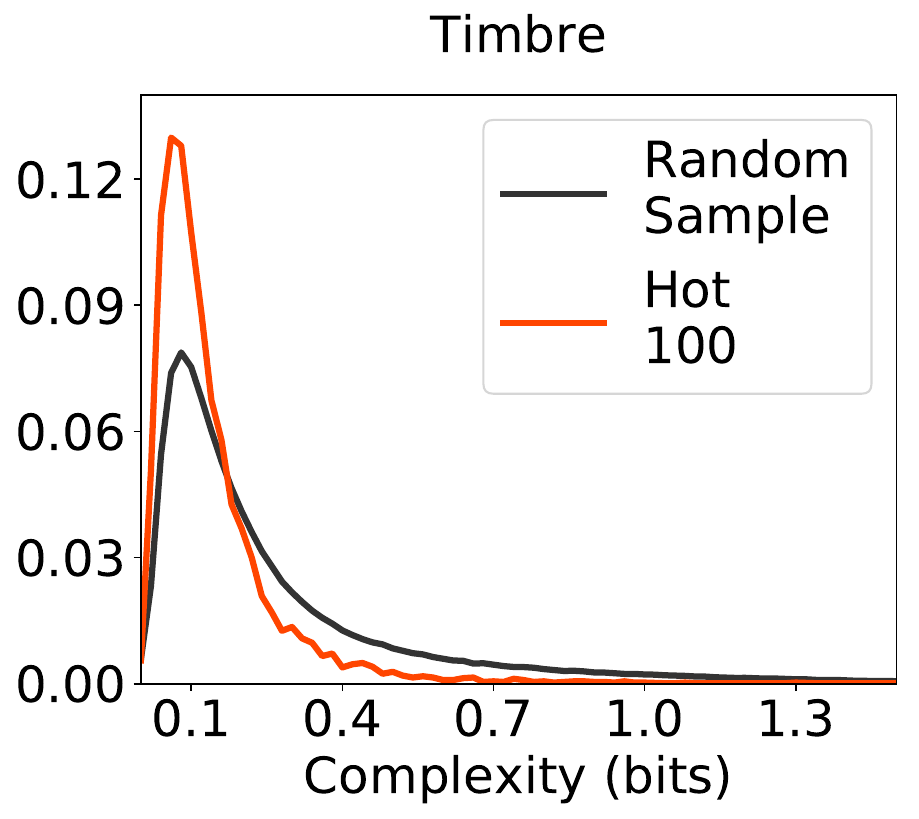}
   \includegraphics[width=.5\columnwidth]{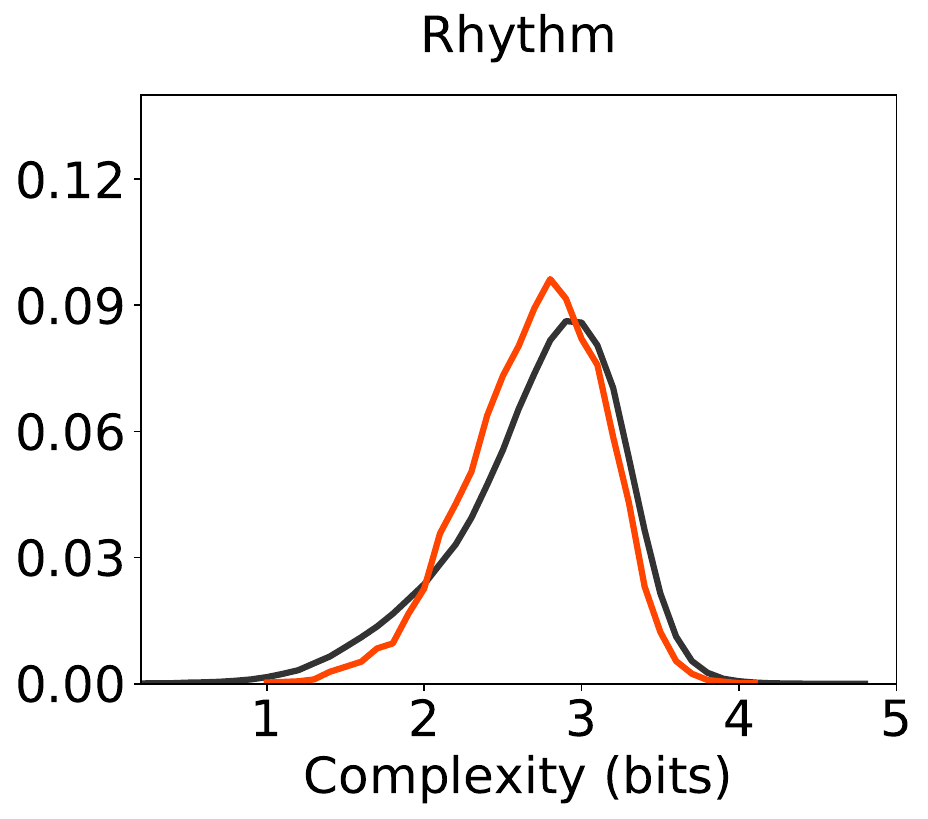}
  
 \includegraphics[width=.51\columnwidth]{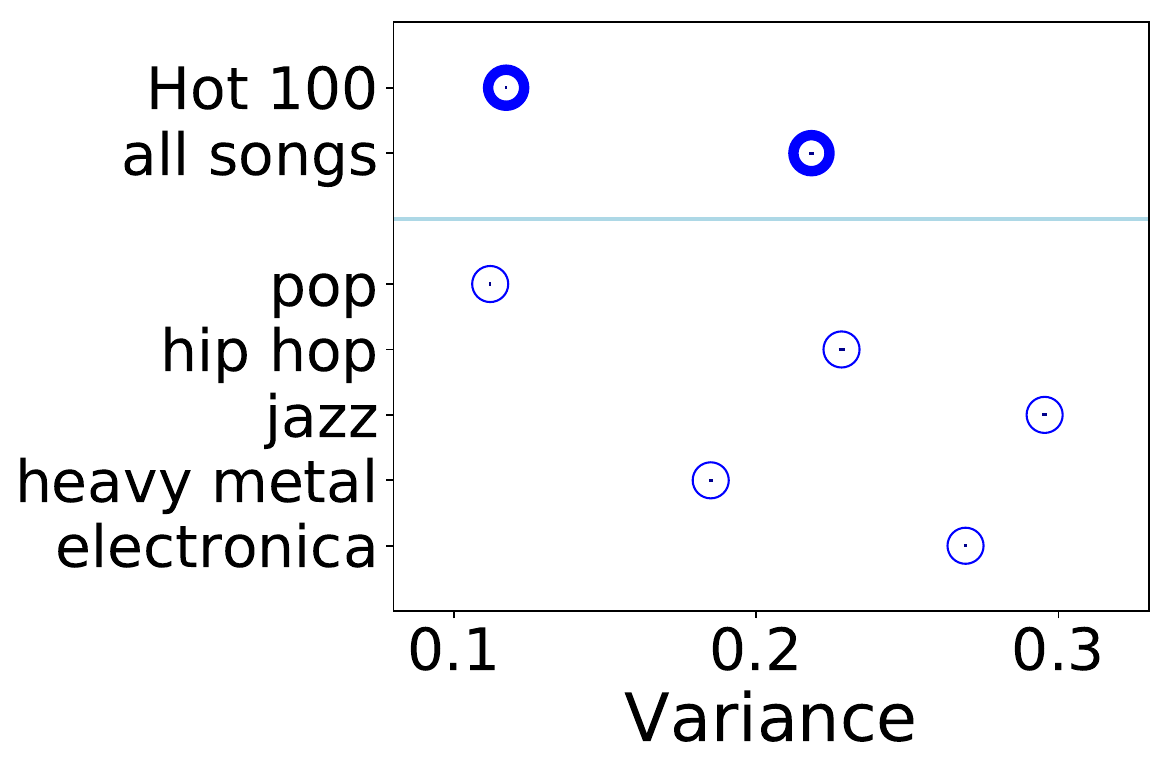}
 \includegraphics[width=.51\columnwidth]{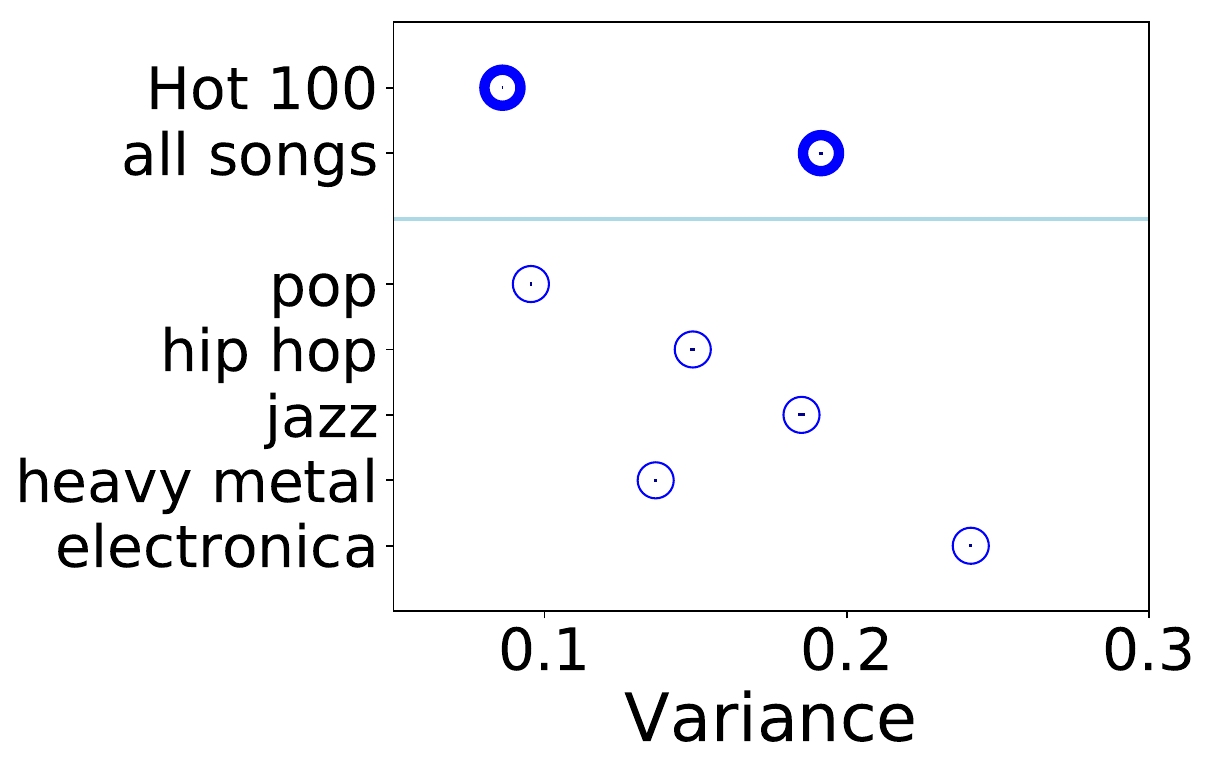}
 \includegraphics[width=.51\columnwidth]{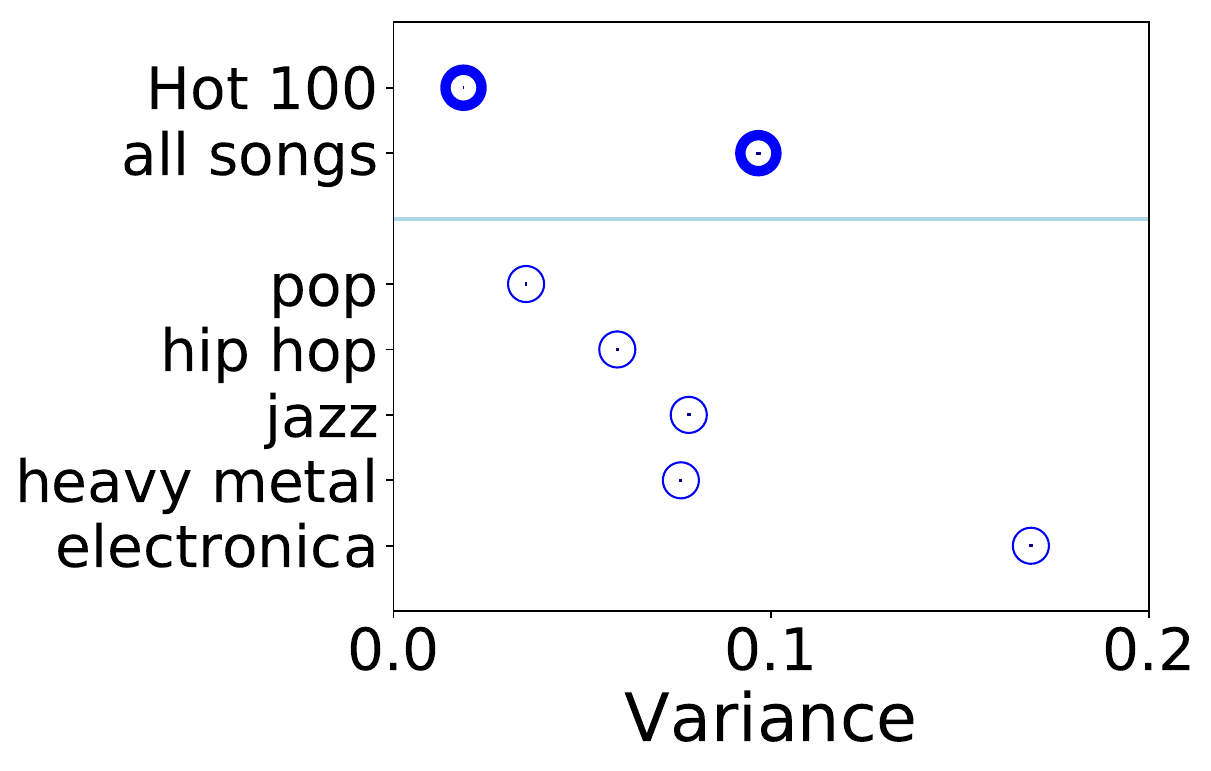}
 \includegraphics[width=.51\columnwidth]{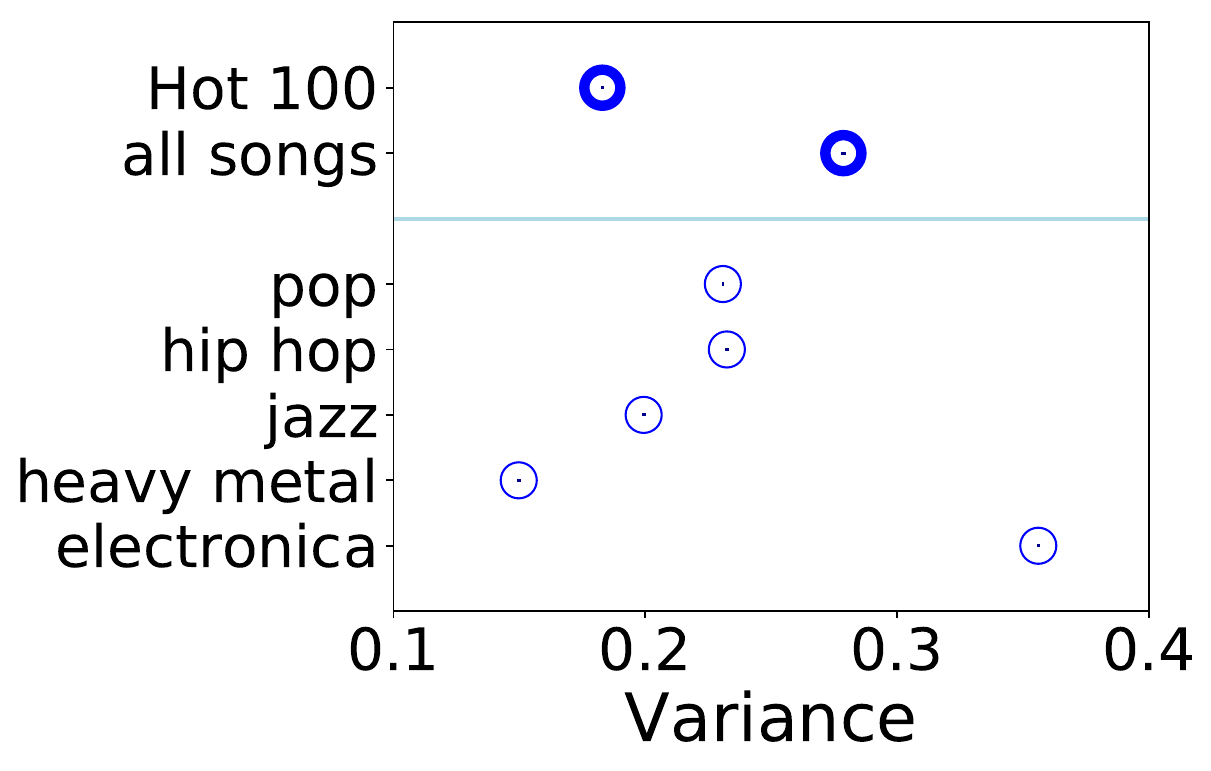}
 \caption{Feature complexities and variances. Complexity distributions are shown (in bins of 0.1 bits, except for timbre which is in bins of 0.02 bits).  The variance plots include 95\% confidence intervals in black (although confidence intervals are smaller than the symbol and not visible), based on 1000 bootstrap samples of 1000 songs from the respective genre.
 }
 \label{fig:complexity_bins}
\end{figure*}

The complexity distribution of songs is approximately bell-shaped, although timbre is skewed towards zero complexity --- unlike the other features, timbre becomes easily predictable after only one previous codeword.  The distributions of the Hot 100 songs are similar, although the Hot 100 tends to exhibit statistically lower complexity in pitch and timbre and higher loudness complexity, compared to 95\% confidence intervals of 1,000 bootstrap random samples of the same size (see Fig.~\ref{fig:complexity_bins}).  Furthermore, we found that the variances of the Hot 100 complexity values are smaller than for other songs (based on 95\% confidence intervals of 1,000 bootstrap samples from the Hot 100 compared to 1,000 bootstrap samples from the overall distribution); thus, the popular songs tend to be located in a narrower range near the mean across pitch, loudness, and rhythm complexity. This result supports the theory for an inverted U-shaped curve where global popularity is maximized by medium complexity.

\subsection{Complexity Across Time}\label{subsec:overtime}

\begin{figure*}
 \centerline{
 \includegraphics[width=1.7\columnwidth]{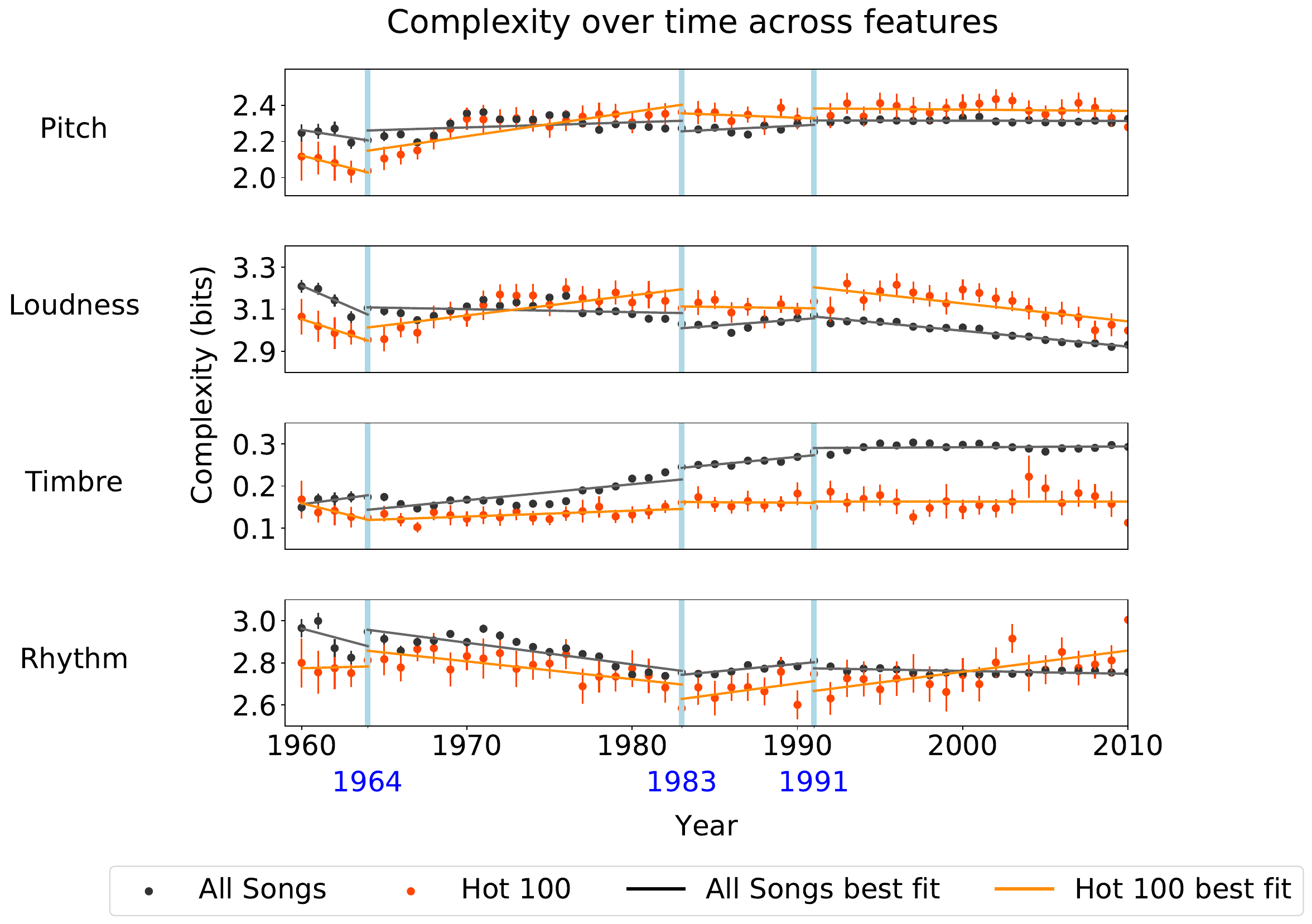}}
 \caption{Average complexity over time across features. Linear trend lines (obtained using OLS linear regression) are shown for each epoch, along with 95\% confidence intervals of the mean.  Light blue lines indicate the musical revolutions found in \cite{Mauch}.}
 \label{fig:complexity_overtime}
\end{figure*}

To examine the evolution of song complexity, we calculate the mean complexity values for each year (for all songs and the Hot 100 songs separately) in Fig.~\ref{fig:complexity_overtime}, which shows several  long-term trends.  Later years mark the appearance of songs with low loudness and rhythm complexity and songs with high timbre complexity, but they were not reflected strongly in the Hot 100 songs.  The low loudness complexity may be due to the trend often called the ``loudness war''~\cite{Serra}, which describes the tendency to produce the entire song to be as loud as possible. Another possible reason may be the emergence of low complexity genres in recent years. For instance, terms associated with low loudness complexity outliers include `grindcore', `hip hop', and `black metal', all of which are relatively newer genres in the dataset. Low rhythm and high timbre complexity may be due to pop or electronic music that contain modern production techniques with many different synthesized textures and strong dance beats.
Terms associated with low rhythm outliers include `tech house', `techno', and `hard trance', while terms associated with high timbre complexity outliers include `tech house', `techno', and `deep house'.

Previous research has indicated that the evolution of Western popular music experienced significant changes during three musical `revolutions' in 1964, 1983, and 1991 \cite{Mauch}.  The first was associated with rock and soul music, the second with disco, new wave, and hard rock, and the third with the emerging popularity of rap music over rock music.  These three revolutions split our period of analysis into three `epochs': 1964-1983, 1983-1991, and 1991-2010.  With this reference frame, we examine our measures of complexity.

If we consider the entire dataset, each aspect of complexity shows a different pattern. The pitch complexity has been more or less stable across the whole period; the loudness complexity has been decreasing overall, although the period from 1983 to 1991 shows a slight increase; the timbre complexity has been steadily increasing and reached a plateau after the 1990s; finally, the rhythm complexity was decreasing through the period from 1964 to 1983, and then stabilized. 

Meanwhile, we find that the temporal evolution of the Hot 100 songs does not follow the overall pattern. The largest difference can be observed in the timbre complexity. While the timbre complexity of the entire dataset has been steadily increasing, it has been almost completely flat for the most popular songs, diverging from the overall trend. This may indicate that the emergence of new genres with high timbre complexity primarily happened for more niche musical tastes.  Pitch and loudness complexity, by contrast, have been higher for popular songs in recent years, while rhythm complexity was lower until the 2000s.

\subsection{Popular Song Similarity}\label{subsec:epochs}

The analysis of complexity over time suggests that modern day popular songs (at least from the 2000s) are more likely to have higher pitch, loudness, and rhythm complexity (and lower timbre complexity) than their less popular contemporaries.  However, while this suggests that popular songs are not simpler than the average song, it does not necessarily indicate whether they sound more or less similar to their popular contemporaries (that is, other Hot 100 songs that are released in the same year).  A recent report suggests that popular songs are sounding more and more similar to other songs on the charts \cite{pudding}.  In contrast, other research finds that songs that perform well on the charts do not sound too similar to their contemporaries but often have an optimal level of differentiation \cite{askin2017makes}.  To analyze whether popular songs become more similar to their contemporaries over time, we measure the Kullback-Leibler (KL) divergence \cite{kullback1951information} from each song in the Hot 100 to other popular songs that were released in the same year.  KL divergence captures the unexpectedness of a song's codewords given the codewords present in other popular songs and thus indicates the spread of codeword usage per year.  

\begin{figure*}
 \centerline{
 \includegraphics[width=1.5\columnwidth]{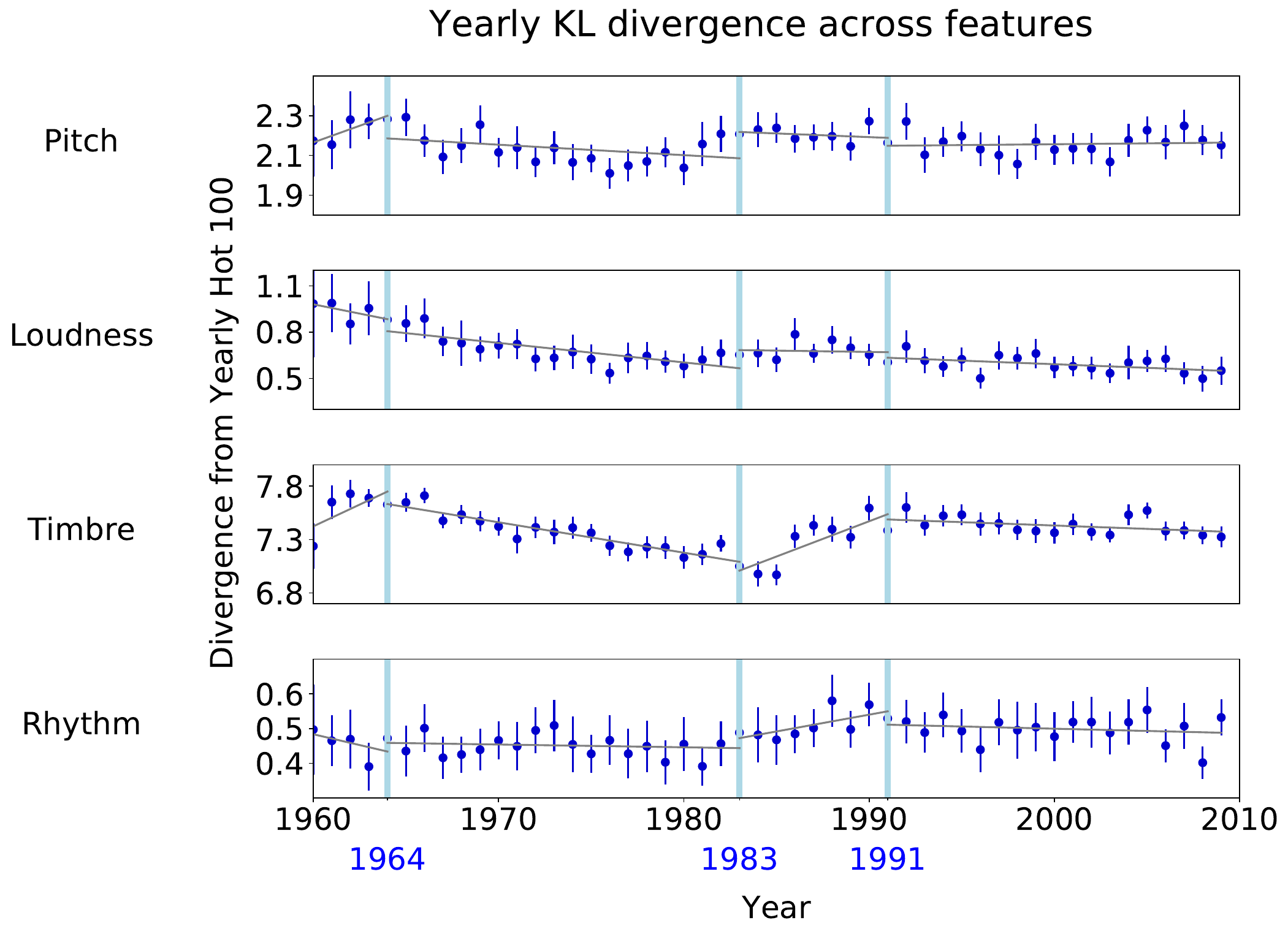}}
 \caption{Yearly KL divergence of Hot 100 songs across features.  Complexity variance, as measured by KL divergence between a song's codewords and codewords from other songs released in that year, shows changing trends at the time of certain musical `revolutions' in 1964, 1983, and 1991 (indicated by light blue lines).  Linear trend lines (obtained using OLS linear regression) are shown for each epoch, along with 95\% confidence intervals of the mean.}
 \label{fig:divergence_overtime}
\end{figure*}

In Fig.~\ref{fig:divergence_overtime}, we show the KL divergence per year for each feature, with each epoch marked. Larger KL divergence in the figure suggests that the songs in that year are more different from their contemporaries as compared to other years. Our measurement shows that the 1964-1983 and 1991-2010 period are similar to each other while 1983-1991 shows a reversing trend. Across features, the KL divergence was either decreasing (songs are more similar to each other) or stable during 1964-1983 and 1991-2010, while 1983-1991 marked either a positive trend reversal or a slow-down of the decreasing trend. These changing trends suggest that musical revolutions are not born equal; some may have spurred diversity among popular songs while some may have homogenized the field. 


Despite these fluctuations, the divergence trend is roughly stable over time for pitch and rhythm, while timbre rebounds after similarity decreases; thus, our findings are consistent with research that songs that perform well on the charts do not sound too similar to their contemporaries but rather maintain a degree of uniqueness which is statistically consistent over time.

\subsection{Complexity Across Genres}\label{subsec:genres}

Let us turn our attention to musical genres and their complexity. As some genres may be characterized by complex harmonic structures or simple, repeated patterns, we expect to see differences across different genres in terms of complexity.  For example, jazz is often considered to have complex patterns whereas dance music may be assumed to use simpler rhythmic patterns.  Our measurement concurs with such speculation, but finds that different subsets of genres may be relatively complex across one or two features but not others.  For instance, electronic and dance styles tend to have high pitch complexity values, whereas jazz and blues have high loudness complexity values.  The highest timbre complexity values belong to electronic genres, although metal also scores highly, but electronic genres have reduced rhythmic complexity which is instead maximal in jazz, progressive and vocal genres. 

\begin{figure}[]
 \centerline{
  \includegraphics[width=\columnwidth]{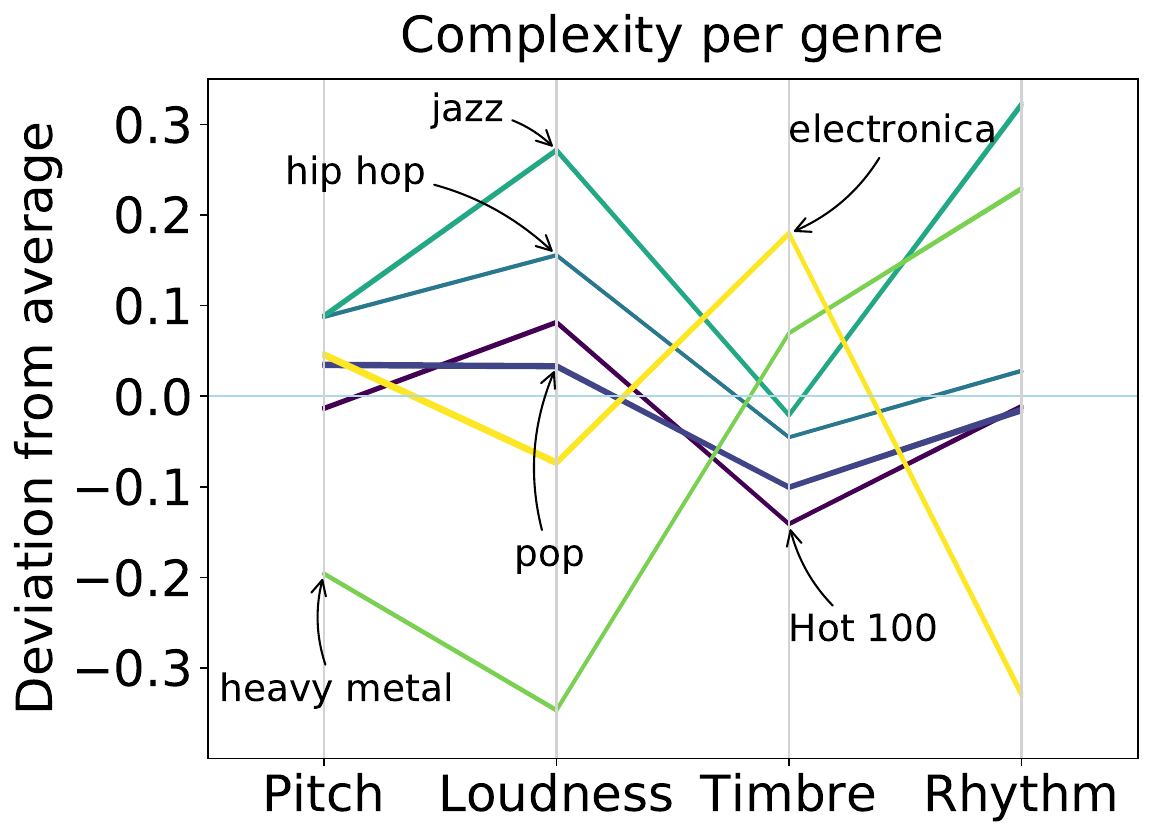}}
 \caption{Complexity of select genres across features.  For each feature, the deviation of the genre complexity mean from the mean of 100 bootstrapped samples of the same size as the genre is shown. Note the similarity between the Hot 100 and the pop genre.}
 \label{fig:complexity_genres}
\end{figure}

We found that a variety of common genres were significantly different from a random sample drawn from the overall distribution in terms of each feature complexity (based on a two-sample Kolmogorov-Smirnoff nonparametric test as well as 95\% confidence intervals of the means of each feature), with the exception that pop was not rhythmically distinct.  Thus each genre seems to have distinctive complexity features that describe its songs: jazz is relatively complex (except in terms of timbre), hip hop has higher than average pitch and loudness complexity, heavy metal has high rhythm complexity but low pitch and loudness complexity, and electronica has high timbre complexity but low rhythm complexity (Fig. \ref{fig:complexity_genres}).

This pattern may be indicative of some trade-offs that listeners make.  If they prefer timbre at the expense of rhythmic complexity, they may prefer electronic genres.  If they prefer pitch and loudness complexity, they may prefer hip hop or jazz.  If they care about timbre and rhythm over pitch and loudness, they may prefer heavy metal.  There is a positive correlation between pitch and loudness complexity (Pearson's r=0.77) across all songs, suggesting that genres tend to have high pitch and loudness complexity (e.g. hip hop, jazz) or low pitch and loudness complexity (e.g. heavy metal).  There is also a negative correlation seen between timbre and rhythm complexity (Pearson's r=-0.55), suggesting that rhythmic complexity decreases with higher timbre complexity (although this is not true for metal genres).  

Interestingly, the Hot 100 is similar to the pop genre in feature means and variances (although statistically different).  Both pop music (whose songs are given no genre-specific term with higher weight than `pop') and the Hot 100 (whose songs are primarily classified as genres other than `pop') have near average values of pitch, loudness, and rhythm complexity, and lower than average values of timbre complexity, while also having smaller variance than the other selected genres (refer to Figures \ref{fig:complexity_bins} and \ref{fig:complexity_genres}).  This may suggest that listeners expect the same from listening to the Hot 100 as they do when listening to music labeled as `pop': mildly surprising songs that do not vary too much in complexity and which are sonically predictable.

One may also expect similar genres to share similar complexity scores.  We used agglomerative clustering on genres represented with over 5000 songs in the dataset (a total of 41 genres), using Euclidean distance between the genre mean complexity scores of each feature, and then used the silhouette coefficient \cite{rousseeuw1987silhouettes,sigaki2018history} to find nine optimal communities.  The result matches intuitive expectations, such that rock genres are grouped together as are electronic genres; interestingly jazz is grouped with hip hop and rap, due to these genres having similar complexity scores. These results suggest that a genre may be defined, to some degree, by its pitch, loudness, timbre, and rhythmic complexity.

\begin{figure*}[]
 \centerline{
 \includegraphics[width=2.1\columnwidth]{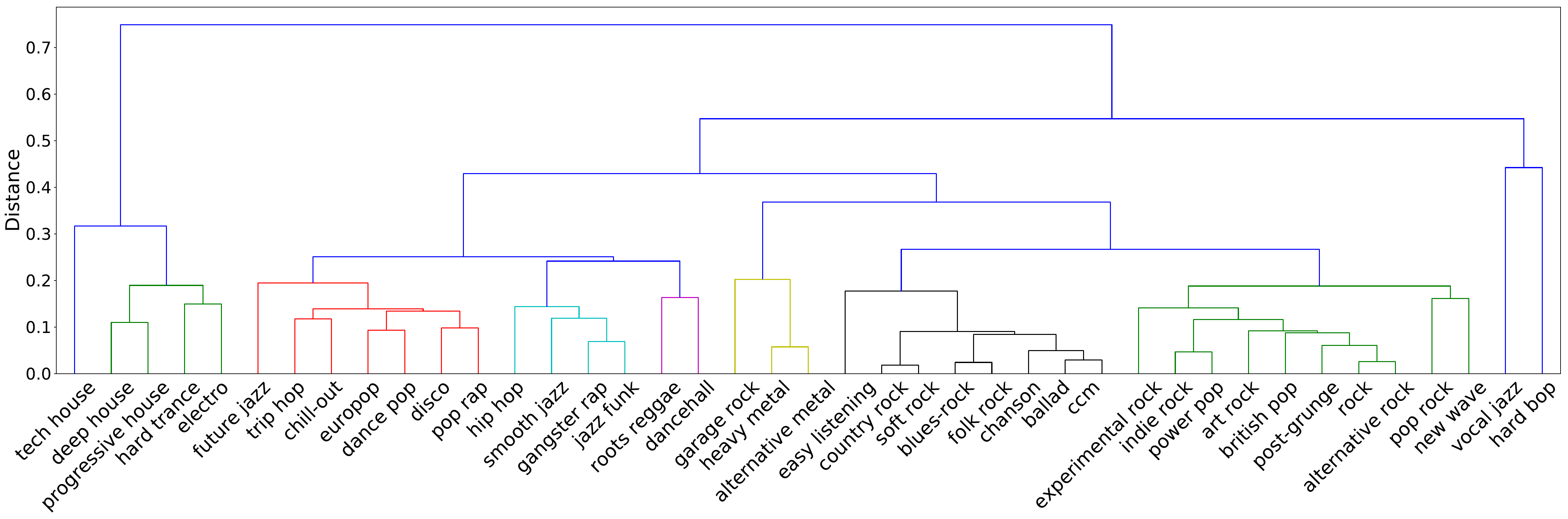}}
 \caption{Dendrogram of top genres.  Clustering is based on the Euclidean distance between complexity features.  Colors indicate different communities.}
 \label{fig:dendrogram}
\end{figure*}

\section{Discussion}\label{sec:discussion}

Understanding and characterizing the complexity of music is an important area of study with both cultural and economic significance. Although music may seem complicated, songs quickly become predictable as you take previous knowledge into account.  This suggests that conditional entropy may be a useful way to characterize musical complexity, although our approach here assumes that the uniform distribution of codewords is the state of maximum uncertainty and expectations are made based on only one previous symbol, which cannot distinguish counts of repeating patterns or phrasing~\cite{Abdallah}.  Our approach is thus intrinsic to the song itself and ignores any \emph{a priori} contextual information.


Using this measure, we find that pitch complexity has been generally stable over the period from 1960 to 2010, while loudness and rhythm complexity have decreased and timbre complexity has increased.  Complexity norms seem to constrain the most popular songs, as those on the Hot 100 are distributed around the overall feature means with small variance in complexity, the exception being timbre which is lower than average.  Indeed, the Hot 100 is similar to songs labeled as `pop', in that pop also has average pitch, loudness, and rhythm complexity and low variance.    

This result provides evidence of a global, inverted U-shaped relationship between popularity and complexity, where popular songs are, on average, the most pleasant to the population. Listeners may expect popular songs to be mildly complex, but not to deviate far from expected timbre or complexity norms.  Complexity of the Hot 100 has in fact been consistent over fifty years in pitch and timbre, while increasing recently in rhythm and decreasing in loudness. 
Similarly, popular songs continue to maintain a consistent level of differentiation from their contemporaries in terms of pitch, timbre, and rhythm.


Certain genres do differ significantly across complexity features, suggesting that they have specific complexity profiles that help define them.  
We hypothesize that certain genres may `make up' for lack of complexity in one area by increased complexity in another.  Perhaps fans of electronic genres prioritize complexity in timbre but rhythmically simple dance beats, or metal fans prioritize rhythmic complexity and high volume at the expense of loudness complexity.

More research needs to be done to fully elucidate the relationship between complexity and musical appreciation.  For example, future research can relate musical complexity to the listening habits of people on a large scale to determine a more fine-grained measure of song popularity.  
The consistency of popular songs over time suggests that, collectively, people tend towards songs that are a certain optimal level of complexity rather than being too simple or complicated.  However, it remains an open question to what degree complexity plays a role in people's cognitive appreciation of music.




\bibliography{music.bib}

%
%
%
%

\end{document}


\maketitle

\section{Filtering the MSD}

The MSD contained songs with bad metadata and segment data, duplicates, and tracks that were talking pieces rather than musical pieces (such as interviews with musicians).  As such, the MSD was filtered to remove any song that met any of the following criteria:

\begin{itemize}
  \item the song had no genre terms
  \item the song had no duration
  \item the song had duration less than three standard deviations below expectation for its genre (where standard deviations where calculated from a random 39100 song sample, directory `A/')
  \item the song title contained the words `interview', `commentary', `introduction', `discuss', `conference', `intro'
\end{itemize}

Additionally, several tracks in the MSD were duplicates of the same song.  Many of these are recognized in the official duplicate list of the MSD \cite{MSDsubset} while others were found by the authors by comparing song and artist names.  Songs were considered duplicates if they shared the same song id or if their song name and artist name matched, subject to the following modifications:

\begin{itemize}
  \item song and artist names were made lowercase
  \item whitespaces were removed
  \item the following characters were removed: " ' , . ? ! ( )
  \item any `\&' was replaced with `and'
  \item any instance of the word `featuring' was replaced with `ft'
\end{itemize}

When considering duplicates, one track was arbitrarily chosen to keep in the filtered set and the rest were discarded.  The resulting filtered dataset contained over 90\% of the original songs.  However not all songs had pitch, loudness, or rhythm data and these songs were kept out of the respective analyses.  Additionally, nearly half of the songs were missing year metadata and so were ignored during overtime analyses.  A list of song ids comprising our dataset is available at https://github.com/tjparmer/music-complexity.

\section{Complexity correlations}

\begin{figure}

{\centering
\includegraphics[width=1.5\columnwidth]{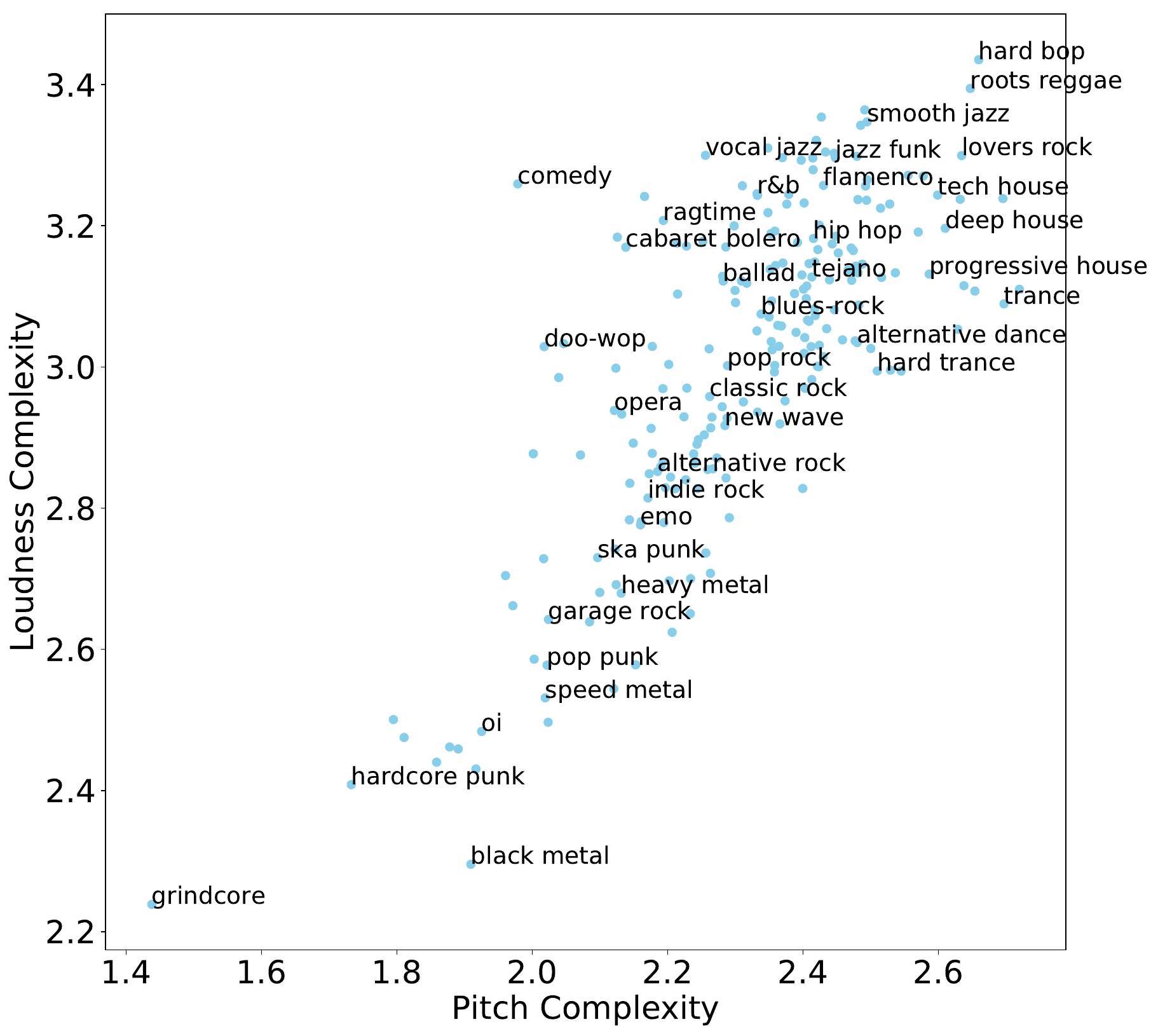}}
\hspace{4cm}
{\centering
\includegraphics[width=1.5\columnwidth]{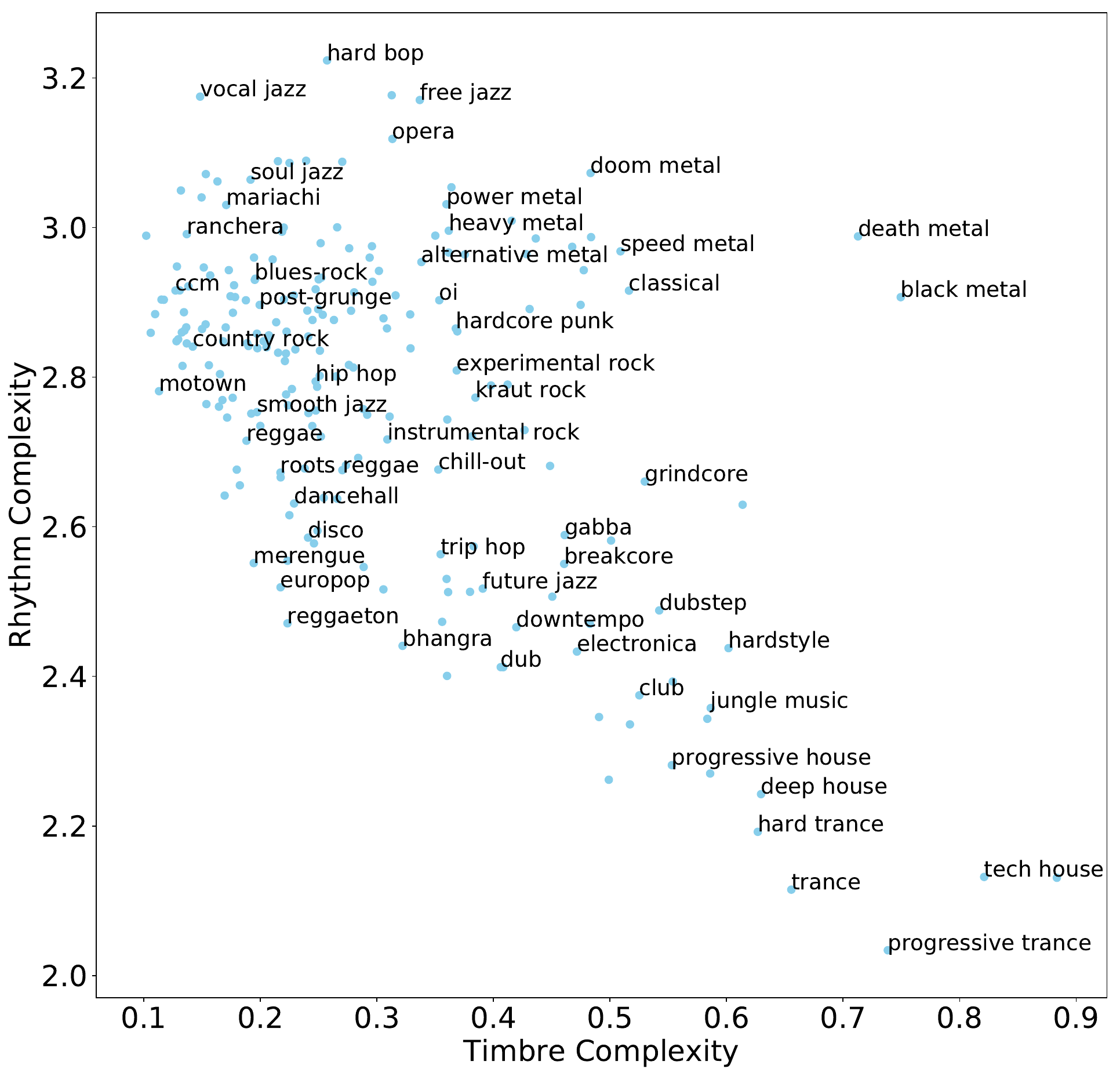}}
\caption{Feature correlations across genres, where each data point represents one genre.  A positive correlation is seen between pitch and loudness complexity (Pearson's r=0.77) and a negative correlation is seen between timbre and rhythm complexity (Pearson's r=-0.55).  Only genres with more than 1000 songs are shown.}
\label{fig:correlations}
\end{figure}

In comparing between every pair of features, two interesting correlations were noted: the positive correlation between pitch and loudness and the negative correlation between timbre and rhythm.  Correlation by genre was calculated using the complexity average of all songs in each genre.  Graphs are shown with genres labeled (refer to Figure \ref{fig:correlations}).

\bibliography{music.bib}